\documentstyle[12pt,aaspp]{article}
\begin{document}
\title{On Turbulent Pressure Confinement of Ultra-Compact HII Regions}

\author{Taoling Xie, Lee G.\ Mundy, Stuart N.\ Vogel}
\affil{Laboratory for Millimeter-wave Astronomy, Department of Astronomy, University of Maryland, College Park, MD 20742; email: tao@astro.umd.edu}
\author{Peter Hofner}
\affil{Universitat zu Koln, I. Physikalisches Institut, Zulpicherstr. 77, 50937 Koln, Germany}
\begin{center}
{\bf Accepted by Astrophysical Journal Letters}
\end{center}
\begin{abstract}
It has been proposed recently that the small size and long lifetime of ultracompact 
(UC) HII
regions could be due to pressure confinement if the thermal pressure of the
ambient gas is higher than previous estimates.
We point out that confinement by thermal pressure alone 
implies emission measures in excess of observed values. 
We show that turbulent pressure, inferred from observed non-thermal velocities, is sufficient to confine UC HII regions and explain their longevity.
We predict an anti-correlation 
between the size of UC HII regions and the velocity dispersion of the ambient 
neutral gas, and show that it is consistent with existing observations.
\end{abstract}

\section{INTRODUCTION}
Ultracompact HII regions (UCHII's) 
are very small ($< 0.2\; pc$) dense regions of ionized gas in molecular clouds
first noted by Ryle \& Downes (1967) and Dreher \& Welch (1981). Deeply embedded, the ionized gas is seen 
only through free-free emission at radio wavelengths.  The importance of this type of object 
for our understanding of the birth of massive stars is well recognized
(Churchwell 1993; Welch 1993; Vogel 1994). A key unsettled question is the
discrepancy between the characteristic age ($\sim 10^{5}\; yr$) of UC HIIs implied by their ubiquity (Wood \& Churchwell 1989, hereafter WC89; Kurtz, Churchwell \& Wood 1994, hereafter KCW94)
and the very short dynamical lifetime ($\sim 10^{3}\; yr$) estimated from 
the sound crossing time.
This disparity implies the presence of physical mechanisms which confine the UC HIIs.
Proposed possibilities include: 
(1) ram pressure due to gravitational infall of ambient gas onto the HII region (cf.\ Reid et al 1981; Dreher \& Welch 1981; Ho \& Haschick 1986),
(2) stellar wind bow shock/ram pressure confinement (cf.\ WC89; van Buren et al 1990),
and (3) 
photo-evaporating disks (Vogel, Genzel \& Palmer 1987; Welch 1993; Hollenbach et al 1994) or photo-evaporating clumps 
(Lizano \& Cant\'{o} 1995; Williams et al 1996), or
champagne-flow/blisters (Tenorio-Tagle 1979; Yorke et al 1983; Forster et al 1990).  

\section{CONFINEMENT BY THERMAL PRESSURE}
De Pree et al (1995) (see also Garc\'{i}a-Segura \& Franco 1996 and Akeson \& Carlstrom 1996) have recently suggested
that the compactness of the UC HIIs can be explained if the density
and temperature in the ambient neutral gas are much higher than commonly assumed. Their basic argument is two-fold. First, the initial Str\"{o}mgren radius $R_{S}$ scales inversely with gas density (e.g. Spitzer 1978) as
\begin{equation}
R_{S}=(\frac{3S_{*}}{4\pi \beta_{2}})^{1/3}n_{0}^{-2/3}=1.99\times 10^{-2}(\frac{S_{*}}{10^{49}\; s^{-1}})^{1/3}(\frac{n_{H2}}{10^{5}\; cm^{-3}})^{-2/3}\; pc,
\end{equation}
where $S_{*}$ is the flux of ionizing photons, $n_{0}=2n_{H_{2}}$ is the initial 
electron density in the ionized gas and $\beta_{2}=2.6\times 10^{-13}\; cm^{3}\;s^{-1}$ is the recombination coefficient. Thus, for a sufficiently high ambient molecular density, the
initial Str\"{o}mgren sphere can be very small. Second, a higher ambient 
gas temperature reduces the amount that the ionized gas expands before
reaching pressure equilibrium with the ambient 
material.

The emission measure of the ionized gas predicted by this simple scenario, however, is significantly larger than the values generally observed. This point can be 
verified easily from the calculations and discussions by De Pree et al 
(1995). For a sample of small UC HIIs, De Pree et al showed that an ambient 
molecular density of $10^{7}\; cm^{-3}$ together with a gas temperature of $\sim 200 \; K$
would suffice to explain the compactness of the UC HIIs. 
They estimated an initial Str\"{o}mgren radius $R_{S}=1.0\times 10^{-3}\; pc$ 
at such high density, corresponding to an initial emission measure 
$EM_{0}=2n_{0}^{2}R_{S}=8\times 10^{11}\; cm^{-6}\; pc$.
Overpressured, the initial ionized sphere expands
outward until it reaches pressure equilibrium with ambient neutral gas. At a 
later time, when the ionized sphere has a radius $R$ and electron density $n_{e}$, the emission measure will drop to 
\begin{equation}
EM=2n_{e}^{2}R=(\frac{R_{S}}{R})^{2}EM_{0}.
\end{equation} 
For $R=0.01 \; pc$, the emission measure is thus expected to be 
$\sim 8\times 10^{9}\; cm^{-6}\;pc$. The observed average emission measures
for spherical UC HII regions with sizes $\sim 0.01\; pc$ are mostly around a few 
$\times 10^{7}$ to a few $\times 10^{8} \; cm^{-6}\;pc$ in the WC89 and KCW94 surveys,
lower than the predicted value by over an order of magnitude, although there clearly are some UC HIIs which have very high emission measures (cf.\ Turner \& Matthews 1984; Garay et al 1993a). 
It seems unlikely that the assumption of optical thinness for the radio emission could have underestimated the emission measure by such a large factor.
Therefore, although the 
gas density around some UC HIIs is indeed higher than $10^{5}\; cm^{-3}$ 
(the density originally adopted by WC89) as recent observations have revealed (cf.\ Huttemeister et al 1993; Akeson \& Carlstrom 1996; Plume et al 1996), a gas density of $n_{H_{2}}= 10^{7}\;cm^{-3}$ 
as adopted by De Pree et al (1995) seem
too high for the material around typical spherical UC HIIs with $R\sim 0.01\; pc$ in WC89 and KCW94 surveys. 
Other physical processes such as swept-up shells or 
dust absorption of Lyman photons may modify the predicted emission measure 
somewhat, but it
is not difficult to see that the former would further increase the emission measure, while the latter 
can only reduce the emission measure by a factor of $(1-x)^{1/3}$, where
$x$ is the fraction of Lyman photons absorbed by dust grains (Franco et al 1990; Churchwell 1993). For $x=0.9$, the emission measure is reduced by a factor of 0.46 (WC89).
While not necessarily irrelevant, dust absorption alone is not likely to be able 
to account for the observed low emission measure for a significant fraction of UC HIIs. We conclude that confinement by thermal pressure requires UC HIIs to have higher emission measures than observed in most cases.
\section{TURBULENT PRESSURE CONFINEMENT}
\subsection{Plausibility}
That the gas pressure in the interstellar medium is not limited to the thermal pressure alone is 
evident from emission line profiles and has recently been brought up in connection with UC HIIs 
by Garc\'{i}a-Segura \& Franco (1996). Ever since the discovery of molecular clouds, it has been 
known that the Doppler-broadened line profiles of all molecular tracers indicate the
presence of significant non-thermal motions, referred to as turbulence, down to the smallest scales 
probed even for low mass molecular structures (cf.\ Falgarone, Puget \& P\'{e}rault 1992).  Since massive star forming regions are
well-known to have stronger turbulence (cf.\ Plume et al 1996), it is likely that turbulence 
exists ubiquitously on the scales of UCHIIs. 
Regardless of whether
the origin is hydrodynamic or hydromagnetic, the presence of turbulence implies a 
significant pressure in addition to the thermal pressure.
Massive star forming regions, where
UC HIIs are found, are particularly turbulent, often with velocity dispersions
well in excess of a couple of kilometers per second (Cesaroni et al 1991; hereafter CWKC; Plume et al 1996). 
The turbulent pressure implied is considerably larger than the
thermal pressure. For example, the turbulent pressure  $P=n_{H_{2}}m_{H_{2}}\sigma_{v}^{2}$ corresponding
to a velocity dispersion of $\sigma_{v}=2\; km\;s^{-1}$ is equivalent to a thermal pressure at an effective kinetic temperature of
$T_{eff}\sim 10^{3}\; K$. In fact, if massive star forming
cores where UC HIIs reside are indeed close to virial equilibrium (CWKC), the 
implied 
molecular gas pressure ought to be comparable to the ram pressure due to free-fall
collapse. Therefore, as long as turbulence exists down to the scales of UCHIIs,
it seems safe to conclude that turbulent pressure could
provide significant opposition to the expansion of the ionized gas, just as would the ram pressure due to gravitational collapse of the molecular envelope (Reid et al 1981). 

The validity of the above hypothesis can be clarified by further considering the two conceptual
phases of UC HII development (Dyson \& Williams 1980; WC89). The first phase, i.e., the establishment of
an initial Str\"{o}mgren sphere, is not affected much by 
the turbulence, so the general results and conclusions in Dyson \& Williams
(1980) and WC89 remain valid, and the Str\"{o}mgren sphere with radius 
determined by Equation (1) will form in a matter of a few years
\footnote{Note that a uniform density is assumed.}.  The temperature of the
newly ionized gas jumps to $T_{H^{+}}\sim 10^{4}\; K$ and the 
number density of the particles in the gas increases by roughly a factor of 4 due to 
dissociation and ionization. This, together with the ram pressure due to stellar 
winds, causes the ionized gas to expand outward at its sound velocity until 
reaching approximate pressure equilibrium with the ambient molecular gas. 
Instead of completely neglecting the pressure due to stellar winds and
turbulence in the ionized gas (Dyson \& Williams 1980; De Pree et al 1995), we 
introduce a factor $\xi$ ($> 1$) to partially take them into account (It can be argued that the ram pressure due to stellar winds may be reflected by
an enhanced actual electron density in a swept-up shell). 
The final average electron density $n_{e}$ in the HII region 
can then be estimated from the following condition (Dyson \& Williams 1980; WC89),
\begin{equation}
2\xi n_{e}kT_{H^{+}}=n_{H_{2}}(m_{H_{2}}\sigma_{v}^{2}+kT_{k})=n_{H_{2}}m_{H_{2}}\delta v_{tot}^{2}/(8ln2),
\end{equation}
where $T_{k}$ is the molecular gas kinetic temperature and $\delta v_{tot}=(8ln2)^{1/2}(\sigma_{v}^{2}+\frac{k}{m_{H_{2}}}T_{k})^{1/2}$
is the FWHM of an optically-thin molecular line tracing the high density gas 
around the UC HIIs. $\delta v_{tot}$ includes both non-thermal and thermal components. If $S_{*}$, the flux of ionizing photons, remains largely unchanged during the 
expansion and the HII region is ionization-bounded, one has 
\begin{equation}
S_{*}(1-x)=\frac{4}{3}\pi R^{3}n_{e}^{2}\beta_{2}=\frac{4}{3}\pi R_{S}^{3}n_{0}^{2}\beta_{2}.
\end{equation}
Equations (3) and (4) give
\begin{equation}
R=R_{S}(\frac{4k\xi T_{H^{+}}}{m_{H_{2}}\sigma_{v}^{2}+kT_{k}})^{2/3}=R_{S}(\frac{32ln2k\xi T_{H^{+}}}{m_{H_{2}}\delta v_{tot}^{2}})^{2/3}.
\end{equation}
So, turbulent pressure can significantly reduce the final equilibrium size of
the UC HII region. If $\sigma_{v}=2 \; km \; s^{-1}$ and $T_{k}\simeq 100\; K$,
we have $R=9.54 \xi^{2/3}R_{S}$ (versus $R=54.3R_{S}$ with thermal pressure alone). For $n_{H_{2}}=8\times 10^{5}\;cm^{-3}$ and $S_{*}=10^{49}\;s^{-1}$ 
(for O6 star), we have $R_{S}=0.005(1-x)^{1/3} \; pc$ and $EM_{0}=2.5\times 10^{10}(1-x)^{1/3}\; cm^{-6}pc$. Thus $EM=(\frac{R_{S}}{R})^{2}EM_{0}=2.8\times 10^{8}\xi^{-4/3} (1-x)^{1/3}\; cm^{-6}pc$ and $R=0.05\xi^{2/3}(1-x)^{1/3}\; pc$. For reasonable dust absorption coefficient $x$ and factor $\xi$, both the predicted size and emission measure are reasonable
in comparison with those inferred from observations (WC89; KCW94), especially considering the expected variation in $n_{H_{2}}$, $S_{*}$ and $\delta v_{tot}$ from region to region. In particular, it appears that the 
initial Str\"{o}mgren sphere expands by a factor of $\sim 10$, say from $\sim 0.01\; pc$ to $\sim 0.1\; pc$, before 
reaching pressure equilibrium with the turbulent ambient molecular gas. The time for the expansion can be roughly 
estimated from the sound crossing time, $\tau_{expansion}\sim \frac{R}{C_{s}}\sim 10^{4}\; yr$, where $C_{s}$ is the sound speed in ionized gas taken as $10\; km\; s^{-1}$.  Since the average age of the UC HIIs observed is close to
$10^{5}\; yr$ (WC89; KCW94), most UC HIIs might 
indeed have had enough time to expand to 
reach pressure equilibrium with their ambient molecular gas (cf.\ Garc\'{i}a-Segura \& Franco 1996).

Note that this scenario, just as the classic picture for HII regions, predicts
that HII regions with larger sizes will have a lower average electron density. This
is consistent with observations by Garay et al (1993b), Churchwell (1993) and KCW94.
Such a size-density correlation is somewhat less obvious for UC HIIs with 
cometary and core-halo morphologies (KCW94; Churchwell 1993). But it is somewhat
difficult to define the average electron density 
for non-spherical UC HIIs and thus the determined densities for these sources
in the literature are likely subject to relatively large systematic errors
(Stan Kurtz 1996, private communication). Further careful observational
studies on this aspect will be useful.
\subsection{Velocity Dispersion-UC HII Size Relation}
Equation (5) indicates that the size of UC HIIs ought to be anti-correlated 
with the total gas pressure or velocity dispersion of the ambient molecular gas 
if the 
gas pressure is indeed responsible for the compactness of UC HIIs.
In the case that $R_{S}$ (i.e., $S_{*}$) does not vary much 
from one region to another, Equation (5) predicts $\delta v_{tot} \propto D_{UC}^{-3/4}$.
However, the surveys by WC89, Garay et al (1993b) and KCW94 imply that $S_{*}$ has a strong 
dependence on $R$ in the sense that larger UCHIIs tend to 
have a larger ionizing flux $S_{*}$ (Churchwell \& Kurtz 1996, private communication)
\footnote{If $S_{*}$ does not vary much from region to region, one expects to see a steep $n_{e}-R$ relation, $n_{e}\propto R^{-1.5}$. In reality, such steep $n_{e}-R$ dependence is not seen (Churchwell 1993; Garay et al 1993b; KCW94)}. 
Assuming that the HIIs are ionization-bounded, the $S_{*}-R$ dependence
can be derived from the
proposed $n_{e}-R$ relation, $n_{e} =n_{e0.1pc} (\frac{R}{0.1\;pc})^{-\alpha}$, where $n_{e0.1pc}$ is the average electron density for UCHIIs with $R=0.1pc$.
Therefore,
$S_{*}(1-x)=\frac{4}{3}\pi R^{3}n_{e}^{2}\beta_{2}\propto R^{3-2\alpha}$, and we have
\begin{equation}
\delta v_{tot} = (\frac{16ln2kT_{H^{+}}}{m_{H_{2}}})^{1/2}(\frac{\xi n_{e0.1pc}}{n_{H_{2}}})^{1/2}(\frac{R}{0.1pc})^{-\alpha/2}.  
\end{equation}
Churchwell (1993) and KCW94 found that $\alpha \sim 0.65$ provides a reasonable 
fit to their data for a sample of spherical and unresolved UCHIIs, while Garay et al (1993b) preferred $\alpha\sim 0.98$ for a larger sample of HIIs with various
morphological types, some of which have relatively large sizes. 

While the size of the UC HIIs can be determined easily from the VLA maps, it is
not entirely trivial to obtain $\delta v_{tot}$ observationally. Specifically,
the following criteria must be considered in choosing molecular lines for 
a reliable determination of $\delta v_{tot}$. First, the molecular line used 
must have a relatively high critical excitation density in order to 
sample the dense gas in the immediate neighborhood of UC HIIs. Second, 
to reduce foreground and background confusion as well as line broadening 
by optical depth effects, the molecular line must be optically thin. Third,
the spatial and spectral resolutions must be adequate.
Existing systematic surveys of molecular gas associated with UC HIIs 
using lower transitions of ammonia and $CO$ and isotopes can be immediately 
ruled out by the above criteria. At this time, it appears that the $C^{34}S$ survey of 8 
UC HIIs by CWKC with the IRAM 30m telescope is the only systematic
data set in the literature from which $\delta v_{tot}$ can be determined
reasonably well for a non-biased sample of UC HIIs.
Among the 3 transitions of $C^{34}S$ observed, the $J=2-1$ data with a 
spatial resolution of $25"$ is preferred
based on considerations of its optical thinness, superior velocity resolution
and reliable detection in all cases. Higher transitions show 
significantly larger linewidths, indicating possible additional broadening 
due to other physical processes such as optical depth effects, and outflows.

Figure 1 shows the data points for all 8 UC HIIs observed in $C^{34}S\; J=2-1$ by 
CWKC in comparison with the expected correlation for pressure-confined UC HII 
regions. The dotted, solid and dashed lines represent Equation (6) with 
$\alpha=0.65$ (Churchwell 1993; KCW94) for $\frac{\xi n_{e0.1pc}}{n_{H_{2}}}=2.55 \times 10^{-2}$, 
$\frac{\xi n_{e0.1pc}}{n_{H_{2}}}=4.04 \times 10^{-2}$ and 
$\frac{\xi n_{e0.1pc}}{n_{H_{2}}}=8.08 \times 10^{-2}$, respectively, i.e. $\delta v_{tot}=(3.16, 2.24, 1.78)D_{UC}^{-0.325}$, where $D_{UC}=2R$. 
Despite a large scatter, an anti-correlation 
between $\delta v_{tot}$ and $D_{UC}$ is evident and in excellent agreement with the theoretical
prediction with $\alpha=0.65$.  In fact, a least-squares-fit to the 8 data points gives 
$\delta v_{tot} (km\; s^{-1}) = (2.45\pm 1.17) D_{UC}^{-0.29\pm 0.07}$ with a correlation coefficient $r=0.87$, which is almost identical 
to the solid line.  Equation (6) with $\alpha=0.98$ (Garay et al 1993b) does not
fit the 8 data points as well, but it is worth noting again that the Garay et al 
sample of UC HIIs contains not only larger HIIs but also non-spherical HIIs for 
which the determined mean electron density is subject to larger uncertainties.
Since $\delta v_{tot}$ does not show any correlation with the distances
to these HII regions,  the possibility that the above 
anti-correlation results from the different physical sizes probed by the same
telescope beam at different distances appears unlikely.
From Churchwell (1993) and KCW94, we have $n_{e0.1pc}\sim 10^{4}\; cm^{-3}$. Therefore the
solid line in Figure 1 implies an average molecular gas density $n_{H_{2}}\sim 2.5\xi \times 10^{5}\; cm^{-3}$.  Given the uncertainties involved, this implied 
mean molecular density is in reasonable 
agreement with the densities derived for these regions (CWKC) and other massive
star forming cores (Plume et al 1996), especially considering the fact that some 
UC HIIs may not reside in the densest molecular gas in massive star forming cores (cf.\ Churchwell 1993; Hofner et al 1996; Xie et al 1996).
Considering the large uncertainty in $D_{UC}$ due to the highly uncertain 
distances to these regions, the consistency between the observations and 
the theoretical prediction for $\delta v_{tot}-R$ relation with $\alpha=0.65$
is surprisingly good.

Finally, we note that $\delta v_{tot}$ may include systematic motions such as collapse and 
outflow or expansion, as observations indeed suggest (cf.\ Ho \& Haschick 1986; Peng 1995; Hofner et al 1996; Shepherd \& Churchwell 1996; Xie et al 1996). The infall of
gas provides additional confining pressure, just like the turbulent motion. 
In the extreme case that $\delta v_{tot}$ is largely dominated by outflowing 
motion, however, there appears an alternative interpretation for the $\delta v_{tot}-D_{UC}$ anti-correlation in the sense that younger and thus smaller UC HIIs
may be associated with larger outflowing motions. We feel that this interpretation is less attractive given the interferometric observations which 
indicate that an UC HII is often just one of a few massive stars in a 
cluster on a considerably larger scale, and the energy input due to an UC HII alone is unlikely to be dominant (cf.\ Wilner, Welch \& Forster 1995; Xie et al 1996).

\section{DISCUSSION}
We also note that it would be naive to expect 
turbulent gas pressure to be the dominant confining mechanism for every UC HII 
region. 
A desirable aspect of the simple turbulent pressure confinement idea is that it 
does not deny the possible role that any other mechanisms may play in addition 
to the turbulent pressure. 

One natural prediction of the turbulent pressure confinement of UC HIIs is
that UC HIIs in relatively quiescent molecular clouds must be rare because they will be short-lived.
Unfortunately, it is difficult to check this 
prediction observationally, because it has long been known that massive
stars tend to form in massive molecular cloud cores where turbulence is known to 
be stronger. However, since even an initially quiescent molecular cloud would become
turbulent as soon as the first generation of stars form and inject kinetic energy
as outflows, one interesting 
speculation is that the first massive stars may develop larger HII regions 
around them and trigger the formation of a later generation of stars
due to the enhanced gas pressure. The stars 
of later generations formed in the pressure environment would have to spend considerably 
longer time embedded in their ambient molecular gas and dust.

Systematic surveys of the molecular gas around a large number of UC HIIs
using optically-thin, high density-sensitive molecular lines with resolution 
on the order of $\sim 10"$ would be highly useful in confirming or 
rejecting the reality of the proposed velocity dispersion-UCHII size relationship and in
clarifying the role that gas pressure plays in the development of HII regions
around newly-formed massive stars. More detailed studies of $n_{e}-R$ or $S_{*}-R$ relation for UCHIIs of various morphologies and sizes will also be very helpful. In particular, it would be highly desirable
if high quality multi-transitional data could be taken for a large number of UC HIIs with a wide range of sizes, which could then be used to determine the density
of the molecular gas as well and thus to check if the UC HIIs are indeed in
rough pressure equilibrium with ambient molecular gas.  

TX thanks Ed Churchwell and Jack Welch for stimulating discussions and 
 in particular for their kind encouragement. He further thanks Neal Evans, Pepe 
Franco, Paul Goldsmith, David Hollenbach, Stan Kurtz, Yuan Peng, Debra Shepherd,
Frank Shu, Frank Wilkin and Qizhou Zhang for useful discussions and Pepe Franco, Paul Goldsmith, Paul Ho, Luis Rodr\'{i}guez and the anonymous referee for reading the manuscript with helpful comments.
This research is partly supported by NSF grant AST9314847 to the
Laboratory for Millimeter-wave Astronomy at the University of Maryland.

\newpage
\begin{center}
{\bf Figure Caption}
\end{center}

Figure 1 shows the theoretically expected linewidth-size relationship versus the
$C^{34}S\; J=2-1$ data from the survey by Cesaroni et al (1991). Despite a large
scatter, the data are remarkably consistent with the theoretical predictions for reasonable parameters, as discussed in the text.
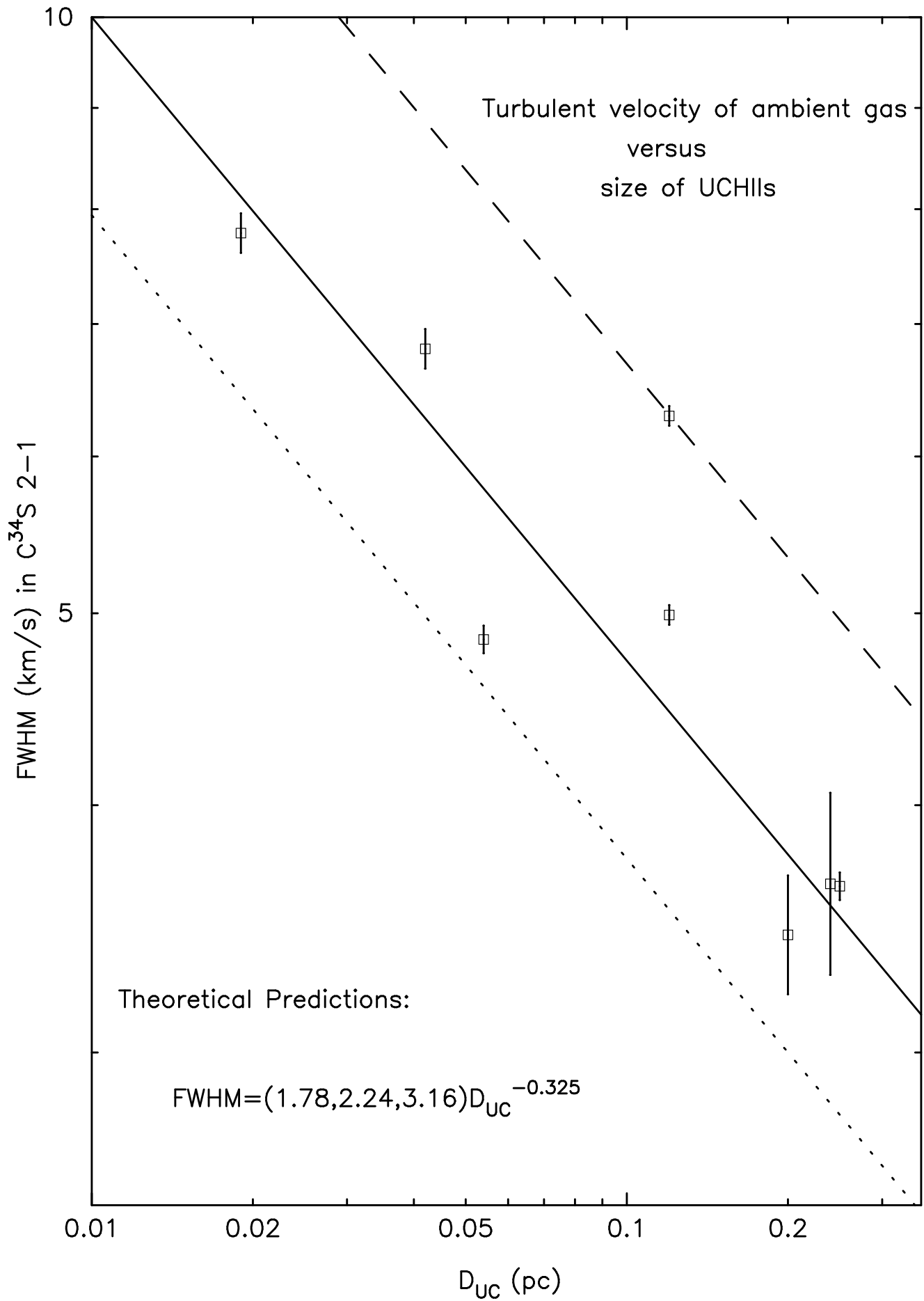
\end{document}